# Slices and Ellipse Geometry

*G. Dattoli*
ENEA – Unità Tecnica Sviluppo di Applicazioni delle Radiazioni,
Laboratorio di Modellistica Matematica, Centro Ricerche Frascati,
Via E. Fermi 45 – 00044 Frascati, Roma (Italy)

*E. Sabia*
ENEA, Unità Tecnica Tecnologie Portici, Centro Ricerche Portici,
P.le E. Fermi 1 - 80055 Portici, Napoli (Italy)

*M. Del Franco*
ENEA Guest

*A. Petralia*
ENEA Fellow

**ABSTRACT**

We discuss the new problems emerging in charged beam transport for SASE FEL dynamics. The optimization of the magnetic transport system for future devices requires new concepts associated with the slice emittance and the slice phase space distribution. We study the problem of electron beam slice matching and guiding in transport devices for SASE FEL emission discussing matching criteria and how the associated design of the electron transport line may affect the FEL output performances. We analyze different matching strategies by studying the relevant effect on the FEL output characteristics.

# I. INTRODUCTION

This paper is devoted to the problem of electron beam slice matching and guiding in transport devices for self amplified spontaneous emission (SASE) free electron laser (FEL) [1]. We will discuss matching criteria and how the associated design of the electron transport line may affect the FEL output performances.

The concept of slice emittance is a by product of the SASE FEL Physics. It is indeed associated with the fact that, in these devices, the combination of mechanisms like gain, slippage and finite coherence length

$$l_c = \frac{\lambda}{4\pi\sqrt{3}\,\rho} \qquad (1)$$

(with $\lambda$ and $\rho$ being the FEL operating wavelength and the Pierce parameter, respectively) determines a kind of local interaction, because the radiation experiences only a portion of the beam, having the dimensions of a coherence length (see Figs. 1). The interaction is therefore sensitive to the longitudinal and transverse characteristics of this "slice", which will be characterized by a specific six dimensional phase space distribution.

In Figure 1a) we have reported an example of a coherent seed, having an rms length of the order of the coherence length undergoing a high gain FEL amplification process, induced by an electron bunch with an rms length $\sigma_z \gg l_c$. In the case of SASE the laser field grows from the noise and therefore, when coherence develops, we have the formation of a number of peaks ($n \cong \frac{\sigma_z}{l_c}$) (identified with the Supermodes [2]). The evolution dynamics is particularly complex and the interplay between these modes during the growth is responsible of the characteristic spiking behaviour [3].

Within certain limits we can consider the evolution of the field generated by an individual spike as due to the characteristics of the corresponding electron bunch slice. The growth of each spike will be therefore determined by various effects associated with the emittance, transverse section and matching condition, energy spread… of the slice determining the local interaction.

In reference [4] we started a preliminary analysis in this direction and studied the effect of the evolution of the Twiss parameters on the laser field evolution, in this paper we will discuss the problem more thoroughly and start with a more complete mathematical analysis, employing concepts from the geometry of conics and therefore we review first some geometrical properties of the ellipses, useful for the analysis of e-bunches phase space slicing.

We either use method of elementary analytical geometry and slightly more advanced techniques employing the formalism of quadratic forms.

We will initially consider the two-dimensional transverse phase space $x, x'$ and define in it the Courant Snyder ellipse [2,5], centred at the origin of the axis (see Figs. 2) and specified by the equation

$$\gamma x^2 + 2\alpha x x' + \beta x'^2 = \varepsilon, \tag{2a}$$

Where $\varepsilon$ is the beam emittance and the coefficients $\gamma, \alpha, \beta$ are the Twiss parameters, linked by the identity

$$\beta\gamma - \alpha^2 = 1 \tag{2b}$$

which ensures the normalization of the phase space distribution

$$\Phi(x, x') = \frac{1}{2\pi\varepsilon} \exp\left[-\frac{1}{2\varepsilon}(\gamma x^2 + 2\alpha x x' + \beta x'^2)\right] \tag{3}$$

Furthermore the emittance and the Twiss coefficients [2,5] define the e-beam rms transverse length, divergence and correlation, according to the relations

$$\begin{aligned}\sigma_x &= \sqrt{\langle x^2\rangle} = \sqrt{\beta\varepsilon}, \\ \sigma_{x'} &= \sqrt{\langle x'^2\rangle} = \sqrt{\gamma\varepsilon}, \\ \sigma_{x,x'} &= \langle x x'\rangle = -\alpha\varepsilon\end{aligned} \tag{4a}$$

where the average is taken on the distribution (3).

The geometrical interpretation of the various quantities, we have mentioned is given in Fig. 2a.

It is worth noting that the two vertical $t$ and horizontal $h$ tangents to the ellipse meet the conic at the points

$$T_1 \equiv (\sqrt{\beta\varepsilon}, -\alpha\sqrt{\frac{\varepsilon}{\beta}}),$$
$$T_2 \equiv (-\alpha\sqrt{\frac{\varepsilon}{\gamma}}, \sqrt{\gamma\varepsilon})$$
(4b)

The geometrical meaning of the correlation $\sigma_{x,x'}$ emerges, therefore, by an inspection to Fig. 2a and is interpreted as the area of the rectangle having as dimensions the coordinates of the tangent points.

The physical role of the $\alpha$ coefficient is that of quantifying the correlation between positions and momenta of the particles in the beam and a further geometrical role is played by its sign, specifying the orientation of the ellipse, which points in the positive direction of the axis for negative values of $\alpha$ and vice versa when it is positive. Ellipses with the same Twiss parameters, but with different emittances, will be said similar (see Fig. 2b)

The angle $\vartheta$ (see Fig. 2c) formed by the ellipse major axis with the positive direction of the $x-$ axis, can be determined by performing the axis rotation

$$x = X\cos(\vartheta) + X'\sin(\vartheta),$$
$$x' = -X\sin(\vartheta) + X'\cos(\vartheta)$$
(5)

and by requiring that the cross terms in $XX'$ vanish, This procedure, which is essentially that of reducing the ellipse to the normal form by imposing that the rotated reference axes coincide with the ellipse axis, yields

$$tg(2\vartheta) = -\frac{2\alpha}{\gamma - \beta}, \gamma > \beta$$
(6).

The above equation as it stands may appear not correct, the Twiss coefficients except $\alpha$ are indeed not dimensionless quantities, and we have $[\beta] = [L], [\gamma] = [L^{-1}]$. The rotation in eq. (5)

which mixes a length ($x$) and dimensionless quantity ($x'$) is not fully legitimate. We can ensure the correctness of the mathematical procedure by defining the rotation with respect to the system of axis $\tilde{x} = \dfrac{x}{\beta_0}, \tilde{x}' = \beta_0 x'$, where $\beta_0$ is a unity with the dimension of a length. Eq. (6) should therefore be written as $tg(2\vartheta) = -\dfrac{2\alpha}{\gamma\beta_0 - \beta\beta_0^{-1}}$. The correctness of eq. (6), from the Physical point of view, is in this way restored, but for practical computation eq. (6) and its consequences can still be used in the form given in the paper.

The use of standard trigonometric relations yields for the angle $\vartheta$ the relations reported below

$$tg(\vartheta) = -sign(\alpha)\left|\dfrac{1+\sqrt{1+r^2}}{r}\right|, \gamma > \beta,$$

$$tg(\vartheta) = -sign(\alpha)\left|\dfrac{1-\sqrt{1+r^2}}{r}\right|, \gamma < \beta \qquad (7).$$

$$r = 2\dfrac{\alpha}{\gamma - \beta}$$

The equation specifying the ellipse major axis in the $x, x'$ plane is, accordingly, given by

$$\begin{aligned} x' &= m\,x, \\ m &= tg(\vartheta) \end{aligned} \qquad (8)$$

We will say that two ellipses are orthogonal if their major axes belong to orthogonal straight lines, accordingly it can be stated that two Courant-Snyder ellipses are orthogonal if their Twiss parameters are such that

$$\begin{aligned} &\gamma, \beta, \alpha \\ &\beta, \gamma, -\alpha \end{aligned} \qquad (9).$$

Examples of orthogonal ellipses are reported in Fig. 3[1].

More in general the mismatch between two ellipses can be quantified through the angle formed by the major axis, as shown in Fig. 3. This angle, in terms of angular coefficient, reads

$$tg(\varphi) = \left|\frac{m_2 - m_1}{1 + m_1 m_2}\right|,$$
$$\varphi = \vartheta_2 - \vartheta_1 \qquad (10).$$

In Figure 4 we show two ellipses forming and the relevant mismatch angle.

We have previously mentioned that the rotation given by eq. (5) allows, with the angle $\vartheta$ specified by eq. (6), to write the equation of the ellipse in the normal form, in this specific case we have

$$\frac{X^2}{A^2} + \frac{X'^2}{B^2} = 1,$$
$$A = \sqrt{\frac{\varepsilon}{\gamma C^2 + 2\alpha C S + \beta S^2}},$$
$$B = \sqrt{\frac{\varepsilon}{\gamma S^2 - 2\alpha C S + \beta S^2}},$$
$$C = \cos(\vartheta), S = \sin(\vartheta) \qquad (11a)$$

and it is easily checked that

$$A B = 1 \qquad (11b).$$

Before closing this section let us consider the two ellipses

---

[1] It is evident that the same considerations hold for the minor axis too, the major and minor axes equations are exchanged in orthogonal ellipses.

$$\gamma_1 x^2 + 2\alpha_1 x x' + \beta_1 x'^2 = \varepsilon_1,$$
$$\gamma_2 x^2 + 2\alpha_2 x x' + \beta_2 x'^2 = \varepsilon_2 \qquad (12)$$

If we multiply both sides of the first equation by $\alpha_2$, the second by $\alpha_1$ and then subtract term by term, we obtain the equation of a conic in normal form, the radical conic, specified by the axes

$$A^2 = \frac{\varepsilon_1 \alpha_2 - \varepsilon_2 \alpha_1}{\alpha_2 \gamma_1 - \alpha_1 \gamma_2},$$
$$B^2 = \frac{\varepsilon_1 \alpha_2 - \varepsilon_2 \alpha_1}{\alpha_2 \beta_1 - \alpha_1 \beta_2} \qquad (13).$$

The two ellipses (1, 2) can have 0, 2, 4 intersection points. In the case of 2 points (bitangent conics) the internal ellipse is the common conic. For no intersections, the common conic is imaginary (it corresponds to both negative sign of $A^2$ and $B^2$) and this aspect of the problem will not be discussed here, because it does not seem to have any relevance to the problems of charged beam transport.

In case of 4 intersection points (radical conic) we may have either an ellipse or a hyperbola, as shown in Figs. 5, according to the sign of $A^2$, $B^2$ the important feature is that it is common to both ellipses, since it passes through the intersection points. If the intersection is an ellipse we call it the "radical ellipse".

The remark on the common conic may sound academic, we will see, in the second part of the paper, that this concept has practical consequence, when referred to charged beam propagation in a transport channel.

In the forthcoming section we will use the formalism of the quadratic forms to recover the previous results from a different perspective.

## II. ELLIPSE GEOMETRY AND QUADRATIC FORMS

The discussion and the results obtained in the previous section can be complemented using a different point of view, based on the formalism of quadratic forms.

The use of two components column vectors and $2 \times 2$ matrices allows to cast the ellipses equation in the form

$$\underline{x}^T \underline{y} = \varepsilon,$$
$$\underline{y} = \hat{D}\underline{x}$$
$$\hat{D} = \begin{pmatrix} \gamma & \alpha \\ \alpha & \beta \end{pmatrix}, \quad (14)$$
$$\underline{x} = \begin{pmatrix} x \\ x' \end{pmatrix}$$
$$\underline{x}^T = (x \quad x')$$

which is essentially a scalar product between the vectors $\underline{x}, \underline{y}$.

On account of eq. (2), the inverse of the $\hat{D}$ matrix reads

$$\hat{D}^{-1} = \begin{pmatrix} \beta & -\alpha \\ -\alpha & \gamma \end{pmatrix} \quad (15)$$

and the ellipse

$$\underline{x}^T \hat{D}^{-1} \underline{x} = \varepsilon, \quad (16)$$

represents the "orthogonal" counterpart of (14).

As we will see in the forthcoming section, the matrix of the quadratic form plays a role analogous to that of the angular coefficient of the straight lines.

The geometry of the ellipse axes can be treated in a different and more elegant way by considering the eigenvectors associated with the matrix $\hat{D}$, specified as it follows[2]

---

[2] The same caveat on the dimensions previously quoted for eq. (6) holds for eq. (17a).

$$\underline{v}_M \equiv \frac{1}{\sqrt{P}} \begin{pmatrix} \frac{1}{2\alpha}(\gamma - \beta - \sqrt{\Delta}) \\ 1 \end{pmatrix},$$

$$\underline{v}_m \equiv \frac{1}{\sqrt{p}} \begin{pmatrix} -\frac{1}{2\alpha}(\gamma - \beta + \sqrt{\Delta}) \\ 1 \end{pmatrix}. \tag{17a}$$

$$\Delta = (\gamma + \beta)^2 - 4$$

where *M, m* refer to major and minor axis respectively and *P, p* are normalization constants.

The angle formed by the vector $\underline{v}_M$ with the *x* axis is given by

$$tg(\vartheta) = -\frac{1}{2\alpha}\left[\gamma - \beta + \sqrt{\Delta}\right] \tag{17b}$$

which is equivalent to the expression given in eq. (6).

The scalar product between two different vectors, individuating the major axes of the relevant ellipses, can be exploited to define the relative orientation of the ellipses. We get indeed

$$\cos(\varphi) = \underline{v}_{M_1}^T \underline{v}_{M_2} \tag{18}$$

when $\varphi = 0$ the two ellipses are aligned, it is also easily checked that $\underline{v}_M^T \underline{v}_m = 0$, as it must be, being the two axes orthogonal.

The alignment between different ellipses in phase space is usually "quantified" by means of the mismatch parameter [6]

$$\zeta = \frac{1}{2}Tr(\hat{D}_1^{-1}\hat{D}_2) = \frac{1}{2}\left[\beta_1\gamma_2 - 2\alpha_1\alpha_2 + \beta_2\gamma_1\right] \geq 1 \tag{19}$$

and perfectly aligned (or parallel) ellipses correspond to $\zeta = 1$, while orthogonal ellipses to $\zeta = \alpha^2$.

The use of this parameter yields a qualitative idea only of how much two ellipses deviate from the perfect alignment, while eq. (18) is a ***measure*** of the mismatch angle.

In the forthcoming section we will analyze this point in a more careful way.

## III. CONDITIONS OF ORTHOGONALITY AND PARALLELISM BETWEEN QUADRATIC FORMS

We have remarked that the matrix of a quadratic form can be viewed as a generalization of the angular coefficient for straight lines. In this section we will develop such a statement more thoroughly.

To this aim we go back to the definition (see eq. (14)) of the vector

$$\underline{y} = \begin{pmatrix} \gamma x + \alpha x' \\ \alpha x + \beta x' \end{pmatrix} \tag{20}$$

and note that the components of the vector belong to two lines in the $x, x'$ plane, with angular coefficients

$$\begin{aligned} m_+ &= -\frac{\gamma}{\alpha}, \\ m_- &= -\frac{\alpha}{\beta} \end{aligned} \tag{21}.$$

We can therefore redefine the vector $\underline{y}$ as

$$\begin{aligned} \underline{y} &= \alpha \hat{M} \begin{pmatrix} x \\ x' \end{pmatrix}, \\ \hat{M} &= \begin{pmatrix} -m_+ & 1 \\ 1 & -\frac{1}{m_-} \end{pmatrix} \end{aligned} \tag{22}.$$

Where $\hat{M}$ will be said the ellipse directory matrix and it plays an interesting role within the present context, which is fully analogous to that of angular coefficient for straight lines.

The inverse of $\hat{M}$ is

$$\hat{M}^{-1} = \alpha^2 \begin{pmatrix} -\dfrac{1}{m_-} & -1 \\ -1 & -m_+ \end{pmatrix} \qquad (23)$$

and it is easily understood that it is the directory matrix associated with the vector

$$\underline{y}^* = \hat{D}^{-1}\underline{x} = \begin{pmatrix} \beta x - \alpha x' \\ -\alpha x + \gamma x' \end{pmatrix} \qquad (24)$$

defining the orthogonal ellipse.

We will accordingly say that two ellipses are parallel or orthogonal, whenever

$$\begin{aligned}\alpha_1 \hat{M}_1 &= \alpha_2 \hat{M}_2, \\ \hat{M}_1 \hat{M}_2 &= \dfrac{1}{\alpha_1 \alpha_2}\hat{1}\end{aligned} \qquad (25).$$

Let us now consider the scalar product

$$\begin{aligned}\underline{y}^{*T}_2 \underline{y}_1 &= (x \; x')\hat{\Pi}\begin{pmatrix} x \\ x' \end{pmatrix}, \\ \hat{\Pi} = \hat{D}_2^{-1}\hat{D}_1 &= \alpha_1\alpha_2 \begin{pmatrix} -1 + m_{+,1}m_{-,2}^{-1} & -(m_{-,2}^{-1} - m_{-,1}^{-1}) \\ (m_{+,1} - m_{+,2}) & -1 + m_{+,2}m_{-,1}^{-1} \end{pmatrix}\end{aligned} \qquad (26)$$

It is easily seen that the mismatch parameter of ref. [6] can be expressed as

$$\zeta = \dfrac{1}{2}Tr\,\hat{\Pi} \qquad (27).$$

On the other side the anti-trace (namely the sum of the off diagonal elements) yields

$$\overline{\zeta} = \dfrac{1}{2}\overline{Tr}(\hat{\Pi}) = \dfrac{1}{2}[(\gamma_2 + \beta_2)\alpha_1 - (\gamma_1 + \beta_1)\alpha_2] \qquad (28)$$

vanishes for parallel ellipses.

## IV NON CENTRAL ELLIPSES

We have introduced so far similar, parallel and orthogonal ellipses, but all the forms we have considered are characterized by the fact that their centre coincides with the axis origin.

More in general if the ellipse is not centred at the origin, its equation can be defined through the equation

$$\gamma x^2 + 2\alpha x x' + \beta x'^2 + 2\delta x + 2\eta x' = \varepsilon', \tag{29a}$$

Or, in matrix form, as

$$\underline{x}^T \hat{D} \underline{x} - 2 \underline{\lambda}^T \underline{x} + \phi = \varepsilon,$$
$$\underline{\lambda} = \begin{pmatrix} -\delta \\ -\eta \end{pmatrix} \tag{29b}.$$

Which can also be rewritten as

$$(\underline{x} - \underline{\chi})^T \hat{D} (\underline{x} - \underline{\chi}) = \varepsilon,$$
$$\varepsilon = \underline{\chi}^T \hat{D} \underline{\chi} + \varepsilon' \tag{29c}$$

where the vector

$$\underline{\chi} = \begin{pmatrix} \xi \\ \rho \end{pmatrix} = \hat{D}^{-1} \underline{\lambda} \tag{30}$$

denotes the ellipse centre coordinates.

This formalism ensures that all the results of the previous sections holds unchanged provided that we consider the appropriate coordinate shift. For example, the equation of the ellipse major axis will determined by

$$x' - \rho = tg(\vartheta)(x - \xi) \tag{31}.$$

In Figure 6 we have reported the example of two similar ellipses having different centres and it is worth stressing that the line connecting the intersection points is specified by the equation

$$x' = -\frac{\delta_1 - \delta_2}{\eta_1 - \eta_2} x + \frac{1}{2} \frac{\varepsilon_1' - \varepsilon_2'}{\eta_1 - \eta_2} \tag{32}$$

If the ellipses are not similar and with different centres we can obtain the common conic by a slight generalization of the procedure leading to eq. 13 and we get

$$(\underline{x} - \underline{\tilde{\chi}})^T \hat{\tilde{D}} (\underline{x} - \underline{\chi}) = \tilde{\varepsilon},$$
$$\tilde{D} = \begin{pmatrix} \tilde{A} & 0 \\ 0 & \tilde{B} \end{pmatrix} \tag{33}.$$

In the forthcoming sections we will see how the notions we have developed so far can be exploited for charged beam transport.

## V. ELLIPSE GEOMETRY AND ELECTRON BEAM TRANSPORT

In the previous sections we have seen that the use of geometrical concepts can be useful to visualize the phase space properties of an e-beam.

In the following we will analyze the usefulness of the just developed formalism, to treat the charged beam transport.

Just to give a preliminary idea of how the previous concepts can be exploited we consider a bunch consisting of two slices, each one specified by the phase space ellipse distributions shown in Fig. 7, passing through an undulator line (see Fig. 8)

It is essentially line of the SPARC experiment [7] and consists of six identical undulator sections.

The ideal beam transport condition is that shown in Fig. 9, which shows the periodic behaviour of the $\beta$ Twiss coefficients, along the line.

It is worth stressing the following "theorem"

***two ellipses with different Twiss parameters describing the phase space conditions of a bunch undergoing the same linear transport system will never become parallel***

It will be therefore never possible to realize a transport system which can realize the optimum matching for two different slices.

In Figures 10,11 we have reported the evolution of two initially orthogonal ellipses along the SPARC undulator line. The figures show that each slices undergoes two completely different stories and the Twiss parameters of one of the slices becomes "out of control". The section and divergence of one slice become wildly large and this effect in the case of FEL SASE operation may determine the conditions for a reduction of the FEL SASE gain, which in turn induces an unacceptable increase of the saturation length.

In the previous figures we have reported the behaviour of the radical ellipse during the transport. Although we have clarified the geometrical meaning of the radical ellipse, its role from the point of view of the beam evolution inside the magnetic line is not clear at all.

We underline that, apart from the slice ellipses, we have essentially three reference ellipses: the projected [1], the radical and the ***best*** ellipse. We introduce the slice brightness according to (see Fig. 12)

$$B_n = \frac{I_n}{\pi^2 \varepsilon_{x_n} \varepsilon_{y_n}}, \tag{35}$$

where $\varepsilon_{\eta_n}$ is the normalized emittance, relevant to the slice $n$ and $I_n$ is the associated current, and define two brightness averaged phase space ellipses, over positive and negative $\alpha$.

The projected slice emittance is that emerging from the brightness averaged phase space distribution (see Fig. 13).

The slice radical ellipse is that emerging from the intersection of the two phase space areas. It is defined as shown in Fig. 13, where we have reported the phase space plots of a sliced bunch. The relevant ellipses are randomly oriented, with an almost equivalent number of cases with $\alpha$ negative and positive.

The best slice is defined as that with Twiss parameters closer to the undulator matching conditions and the smallest emittance.

As is well known the concept of slice emerges from the SASE FEL Physics and brings the non secondary question of defining the most convenient electron beam transport strategy for the laser process.

The problem can be understood as it follows: the FEL-SASE gain length is defined as [8]

$$L_g = \frac{\lambda_u}{4\pi\sqrt{3}\,\rho} \qquad (36)$$

and it is inversely proportional to the Pierce parameter $\rho$, which in turn depends on the current density. Each electron bunch slice will therefore determine a corresponding laser slice, which will be characterized by its own gain length and the slices with shorter gain length (and thus with larger $\rho$) are those exhibiting a shorter saturation length.

Appropriate $\rho$ values require reasonable emittance, reasonable current and reasonable matching of the Twiss coefficients during the transport of the e-beam inside the undulator. If the beam section or its divergence become too large, the associated reduction of the current density determines a corresponding increase of the gain length and eventually create problems to the saturation.

In the following we will consider the case of the SPARC experiment and describe with some detail different matching strategies and the relevant consequences on the SASE FEL dynamics.

The SPARC transfer line is reported in Fig. 14 and in Figs. 15 we have reported the evolution of the $\beta - Twiss$ parameters, inside the line, for the different slices of the bunch, whose characteristics are reported in Table I.

The electrons are then injected inside the undulator transport line (see Figs. 8-9) and the goal is that of avoiding significant oscillations of the slice Twiss parameters. The concept of beam matching becomes rather doubtful, since we cannot adapt the line to provide the minimum of the beta function of each slice at the centre of each undulator section.

A large excursion of the beta function may be responsible for a significant reduction of the corresponding current density, we expect therefore that this causes an increase of the saturation length and/or a reduction of the output laser power.

We will analyze the consequence of the matching strategy on the SASE FEL evolution by studying three different possibilities by matching on

a) The best slice, namely the ellipse with Twiss parameters closer to the undulator matching conditions and the smaller emittance

b) The projected ellipse

c) The radical ellipse

The slice Twiss parameters transport is reported in Figs. 16, where we have shown the evolution along the undulator $z$ axis, for the assumptions a-c). The parameters of the best slice in our simulation are reported in Table I for the slice of index n=6. It is evident that the matching on the radical ellipses provides the most convenient choice since it avoids large oscillations of the Twiss parameters even for the completely mismatched cases.

The consequences of the previous choices on the evolution of the laser field will be discussed in the forthcoming section, where we will follow the evolution effective, picture in which we will study the interplay with the evolution of the Twiss parameters.

## VI. GAIN PROCESS FOR SLICE, PROJECTED AND RADICAL EMITTANCE

In the following we will model the problem of the laser field evolution inside the undulator, by making an approach combining the semi-analytical procedure developed in ref. [8] and by introducing the effect of the Twiss parameter excursion through the gain length, which depending on the current density is sensitive to the variation of the beta Twiss parameters, in turn linked to the transverse beam sections. We will use therefore the logistic equation

$$P(z_b, z) = \sum_n h_n(z_b) \cdot P_0 \cdot \frac{A_n(z)}{1 + \frac{P_0}{P_{F,n}(z)} \cdot (A_n(z) - 1)} \tag{37}$$

with

$$A_n(z) = \frac{1}{9}\left(3 + 2\cosh\left(\frac{z}{L_{g,n}(z)}\right) + 4\cos\left(\frac{\sqrt{3}}{2}\frac{z}{L_{g,n}(z)}\right)\cdot\cosh\left(\frac{z}{2L_{g,n}(z)}\right)\right),$$
$$L_{g,n}(z) = \frac{\lambda_u}{4\pi\sqrt{3}\rho_n(z)}$$
(38).

Where $z$ is the longitudinal coordinate, along the undulator axis, $L_{g,n}(z)$ the local gain length and $\rho_n(z)$ is "local" Pierce parameter.

By recalling indeed that

$$\rho \propto J^{\frac{1}{3}}, J \propto (\sigma_x \sigma_y)^{-1}$$
(39)

Where $\sigma_{x,y}$ are the e-bunch transverse coordinate, which are expressible as

$$\sigma_\eta = \sqrt{\beta_\eta \varepsilon_\eta}, \eta = x, y$$
(40)

The local Pierce parameter is therefore that which takes into account the values of the beta parameter at each point of the undulator.

The index $n$ denotes the slice index, $h_n(z_b)$ the slice distribution along the bunch coordinate $z_b$, $P_0$ is the input seed (assumed to be the same for all the slices) and $P_{F,n}(z)$ is the local saturated power density.

The model does not include diffraction, energy spread and inhomogeneous broadening . We have neglected any information on the phase evolution and included only the part relevant to the intensity evolution, the model does not account for possible spiking behaviour, since it deals with the effects of the transverse phase space distribution on the SASE field power growth.

In Figures 17 we have reported the power growth evolution, associated with the best slice, under the a-c conditions of the previous section.

It is evident that the matching on the radical emittance seems to provide the best performances, in terms of power saturation length.

The consequences of the matching for the other slices are shown in Figs. 18 where we have reported the evolution of 7 different slices (see Table II) exhibiting different characteristics.

The results shown in the figure are fairly interesting. Although it is confirmed that the overall best performances seems to be provided by the matching on the radical ellipse, the matching on the projected emittance is that which keeps the slice evolution closer even for that with the worst matching characteristics.

Finally in Fig. 19 we have reported the sum of the power of each slice vs. the undulator length for different matching conditions. Albeit we have just provided a naïve overlapping without taking into account the spiking contributions the plots indicate again that different matching strategies, yields different performances in terms of output power and saturation length.

The analysis we have developed assumes a kind of squared bunch, we have indeed assumed that all the slice have the same current, therefore once neglected the transverse phase space evolution effects all the slices should exhibit the same evolution curve.

The inclusion of the effects due to the dependence of the current on the bunch coordinate can be included straightforward and will not be discussed here, because they do not add further interesting information.

A further important effect not mentioned here is that of slippage. The optical fields having a larger velocity than the electrons slips over the electron bunch and explores different regions of the bunch during the evolution inside the undulator.

We have already shown [4] that the inclusion of this effect in the present model, is feasible without significant problems. In a forthcoming investigation we will generalize the present analysis by a systematic investigation of the slippage effect within the framework of the present model and by including the study of the linear harmonic generation.

## VII. CONCLUDING REMARKS

In the previous section we have considered the transport of slices all having phase space distribution with the same centre. Non central distribution may be due to the effect of a chromatism induced on the electron bunch different slices, having different energies,

In Figure 20 we report the slice phase space ellipses, with random centres, this may be the source of further complications as the increase of the effective beam emittance due to the centroid scatter. We will include the effect of non central emittance by exploiting the procedure suggested in refs. [4,9]. We will define indeed the following centroid r.m.s. position, divergence and correlation

$$\sigma_\xi^2 = \langle \xi^2 \rangle - \langle \xi \rangle^2,$$
$$\sigma_{\xi'}^2 = \langle \xi'^2 \rangle - \langle \xi' \rangle^2,$$
$$\sigma_{\xi,\xi'} = \langle \xi \xi' \rangle - \langle \xi \rangle \langle \xi' \rangle,$$
$$\langle a \rangle = \frac{1}{m} \sum_{n=1}^{m} a_n$$

(39)

and the associated emittance and Twiss coefficients, namely

$$\varepsilon_\xi = \sqrt{\sigma_\xi^2 \sigma_{\xi'}^2 - \sigma_{\xi,\xi'}^2},$$
$$\beta_\xi = \frac{\sigma_\xi^2}{\varepsilon_\xi}, \gamma_\xi = \frac{\sigma_{\xi'}^2}{\varepsilon_\xi}, \alpha_\xi = \frac{\sigma_{\xi,\xi'}}{\varepsilon_\xi}$$

(40).

The phase space distribution generated by a random sorting of different ellipses centres is shown in Fig. 20.

According to the previous picture the centroid scattering produces a kind of diffusion which contributes to the e-beam emittances and it can be accounted for by redefining the slice phase space distribution according to the following convolution

$$\widetilde{f}_n(x,x') = \int_{-\infty}^{\infty} f_c(\xi,\xi') f_n(x-\xi, x'-\xi') d\xi d\xi'$$

(41)

where $f_c(x,x')$ is the scattered centroid equivalent phase-space distribution and $f_n(x,x')$ is the slice phase space distribution. An example of the effect of such a convolution is shown in Fig. 20. Within this framework we can treat the problem of the slice propagation inside the transfer and the undulator transport line, using the same procedure as before.

Each slice phase space distribution is then specified by means of a convolution on the (41). We can define the associated phase space distribution and specify the projected distribution as a convolution between the two having the momenta given by eqs. (29) and (9). In this way the centroids contribute to the projected emittance in a kind of diffusive way. The centroid spreading can be due to various mechanisms, which will be carefully discussed in a forthcoming investigation.

Here we want to stress the relevance of the slice energy spread. If any slice is characterized by a longitudinal energy distribution of the type

$$\varphi_n(z,\varepsilon) = \frac{1}{2\pi \Sigma_{\varepsilon,n}} \exp\left(-\frac{\beta_{\varepsilon,n}\varepsilon^2 + \gamma_{\varepsilon,n}z^2}{2\Sigma_{\varepsilon,n}}\right) \tag{31}$$

with $\Sigma_{\varepsilon,n}$ being the longitudinal emittance and with

$$\varepsilon = \frac{\gamma - \gamma_0}{\gamma_0} \tag{32}$$

being the relative energy. The single slice energy spread and bunch length can therefore be defined as

$$\begin{aligned}\sigma_{\varepsilon,n} &= \sqrt{\gamma_{\varepsilon,n}\Sigma_{\varepsilon,n}}, \\ \sigma_{z,n} &= \sqrt{\beta_{\varepsilon,n}\Sigma_{\varepsilon,n}}\end{aligned} \tag{33}.$$

The "chromatic" structure of the packet will reflect itself into the magnification of the chromatic effects inside transport elements like quadrupoles and solenoids, any individual slice will indeed be affected in a different way by the energy dependent part of the transport element.

An example may be provided by a solenoid misalignments [10], causing a vertical component of the magnetic field, which will induce a coupling of the motion in the $x-z$ plane, which will be characterized by centroid shift, which can be expressed as

$$\xi_n \propto \Lambda^{-1} \sigma_{\varepsilon,n},$$
$$\Lambda = \frac{m_0 \gamma_0 c}{e\, \delta B} \tag{34}$$

The distribution given in eq. (31) does not contain any energy phase correlation term, whose effect will be discussed elsewhere along with its important role in the physics of FEL.

## ACKNOWLEDGMENTS

The Authors express their sincere appreciation to Drs. F. Ciocci, L. Giannessi and C. Ronsivalle for a number of enlightening discussions on practical aspects of charged beam transport.

# FIGURE CAPTIONS

Fig. 1 -   Local interaction and slice emittance

Fig. 2 -   a) The Courant Snyder ellipse and the relevant geometrical interpretation; b) Similar ellipses; c) Definition of ellipse's angles with the x axes, it can be positive ($\theta_1$) or negative ($\theta_2$).

Fig. 3 -   Example of orthogonal ellipses with equal and different emittances.

Fig. 4 -   Ellipses forming an angle of $\frac{\pi}{6}$

Fig. 5 -   Ellipses intersection and common conic.

Fig. 6 -   Similar non central ellipses and radical axis.

Fig. 7 -   Orthogonal ellipses at the entrance of the undulator line.

Fig. 8 -   Undulator line at SPARC.

Fig. 9 -   Behaviour of the beam beta Twiss coefficients for the ideal transport in the SPARC undulator line, $\beta_x$ continuous line, $\beta_y$ dotted line.

Fig. 10 -  Transport of two different slice along the undulator. The figures report the ellipses phase space with the radical ellipse (black line) at different position in the undulator: a) z=0, b) z=Lu/4, c) z=3Lu/4, d) z=Lu.

Fig. 11 -  (a) Angle between two slices, (b) Mismatch parameter and (c) Anti-trace (eq. (28)): evolution along the undulator longitudinal coordinate z.

Fig. 12 -  Slice current and associated brightness of the sample slice distribution used in the simulation.

Fig. 13 -  Courant Snyder ellipse in the phase space. a) ellipses of different slices and the resultant projected emittance (green line); b) the two mean ellipse with positive (red) and negative (blue) angle respect to the x-axes, the resultant radical ellipse is the black line; c) radical ellipse (black) versus projected ellipse (green).

Fig. 14 -  SPARC Transfer line from Linac exit (on the left) to the Undulator entrance (right). In red we see the two triplets of quadrupoles.

Fig. 15 -  Radial (a) and vertical (b) beta-Twiss parameters in the SPARC transfer line for the n slices. The green bars indicate the position of the magnetic elements along the line.

Fig. 16 -  Radial slice Twiss β parameter. a) the matching to the undulator is done in such a way that the slice with best Twiss parameters and smallest (n=6) emittance is well transported; b) the matching is done by optimize the transfer line for the projected emittance; c) the optimization is done for the radical ellipse.

Fig. 17 -  Power growth of the best slice (n=6). a) by matching on the best slice; b) by matching on the projected emittance; c) by matching on the radical emittance.

Fig. 18 -  Power growth of sample slices (index 1,4,5,6,7,8,10) along the undulator a) by matching on the best slice; b) by matching on the projected emittance; c) by matching on the radical emittance.

Fig. 19 -  Power growth along the undulator line of the sum of all slices for different matching conditions: a) on best slice, b) on projected emittance, c) on radical ellipse .

Fig. 20 -  The contribution to the increase of the projected emittance of the phase space centre mismatches of the individual slices (the slash contour in the plot on the right refers to projected emittance for slices having all the same phase space centre).

TABLE I. Slice parameter of the sample slice distribution. The parameters are the ones at the exit of the Linac.

| Slice index | I (A) | $\varepsilon_x$ (mm·mrad·$\gamma_0^{-1}$) | $\alpha_x$ | $\beta_x$ (m) | $\varepsilon_y$ (mm·mrad·$\gamma_0^{-1}$) | $\alpha_y$ | $\beta_y$ (m) | Brigth. ($10^{17}$ A/m$^2$) | J ($10^8$ A/m$^2$) |
|---|---|---|---|---|---|---|---|---|---|
| 0 | 20 | 1.088 | -10.709 | 18.553 | 1.008 | -1.035 | 16.802 | 5.0019 | 1.0107 |
| 1 | 20 | 3.282 | -4.695 | 17.791 | 4.625 | 6.149 | 16.142 | 0.3614 | 0.2830 |
| 2 | 20 | 1.01 | 2.825 | 16.744 | 1.151 | -1.457 | 16.088 | 4.7187 | 1.0560 |
| 3 | 20 | 1.029 | -1.5 | 16.147 | 1.102 | -1.621 | 15.919 | 4.8376 | 1.0946 |
| 4 | 20 | 1.071 | -1.056 | 15.463 | 1.155 | 1.199 | 15.891 | 4.4346 | 1.0719 |
| 5 | 20 | 2.09 | 0.921 | 15.438 | 3.183 | 1.122 | 15.633 | 0.8246 | 0.4664 |
| 6 | 20 | 1.084 | 1 | 15 | 1.164 | 1 | 15 | 4.3475 | 1.1091 |
| 7 | 20 | 1.056 | 1.88 | 15.469 | 1.124 | 1.274 | 15.681 | 4.6216 | 1.1013 |
| 8 | 20 | 1.058 | 1.298 | 15.782 | 1.116 | 1.589 | 15.732 | 4.6459 | 1.0914 |
| 9 | 20 | 4.009 | 2.85 | 15.914 | 2.096 | -1.873 | 16.538 | 0.6528 | 0.3974 |
| 10 | 20 | 10.126 | -4.892 | 16.783 | 1.4 | 5.817 | 16.61 | 0.3870 | 0.2973 |
| 11 | 20 | 2.215 | -5.561 | 16.24 | 3.3 | -8.887 | 16.944 | 0.7505 | 0.4167 |
| 12 | 20 | 0.575 | 7.986 | 17.287 | 1.187 | -10.162 | 17.486 | 8.0372 | 1.3010 |

TABLE II. Parameters of the sample slices of the sample distribution. The best parameters are marked in boldface.

| Slice index | I (A) | $\varepsilon_x$ (mm·mrad·$\gamma_0^{-1}$) | $\alpha_x$ | $\beta_x$ (m) | $\varepsilon_y$ (mm·mrad·$\gamma_0^{-1}$) | $\alpha_y$ | $\beta_y$ (m) | Brigth. ($10^{17}$ A/m$^2$) | J ($10^8$ A/m$^2$) |
|---|---|---|---|---|---|---|---|---|---|
| 1 | 20 | 3.282 | -4.695 | 17.791 | 4.625 | 6.149 | 16.142 | 0.3614 | 0.2830 |
| 4 | 20 | **1.071** | -1.056 | **15.463** | **1.155** | **1.199** | 15.891 | **4.4346** | **1.0719** |
| 5 | 20 | 2.09 | 0.921 | **15.438** | 3.183 | **1.122** | 15.633 | 0.8246 | 0.4664 |
| **6** | 20 | **1.084** | 1 | 15 | 1.164 | 1 | 15 | 4.3475 | 1.1091 |
| 7 | 20 | **1.056** | 1.88 | 15.469 | 1.124 | 1.274 | 15.681 | 4.6216 | 1.1013 |
| **8** | 20 | **1.058** | 1.298 | 15.782 | 1.116 | 1.589 | 15.732 | 4.6459 | 1.0914 |
| 10 | 20 | 10.126 | -4.892 | 16.783 | **1.4** | 5.817 | 16.61 | 0.3870 | 0.2973 |

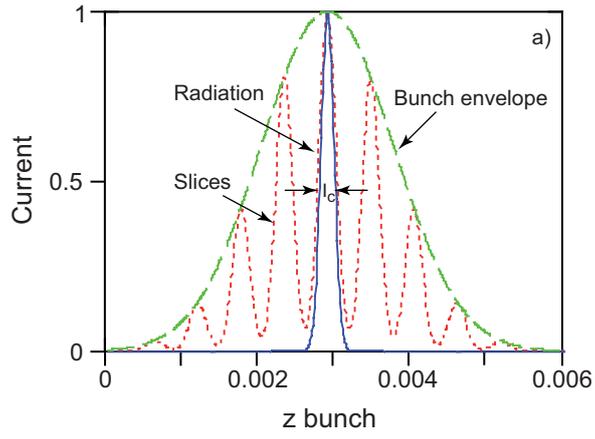

Fig. 1a

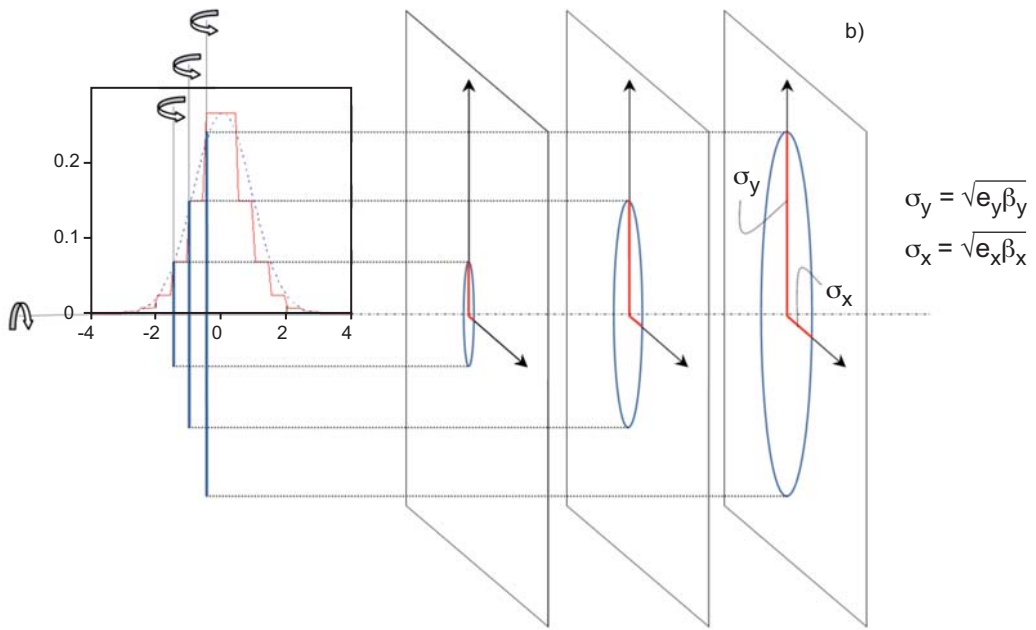

Fig. 1b

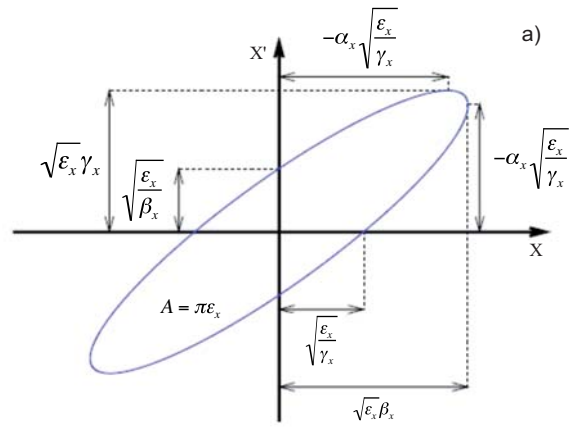

Fig. 2a

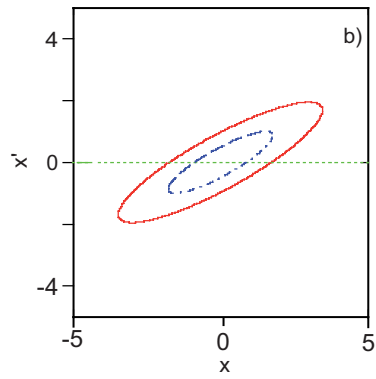

Fig. 2b

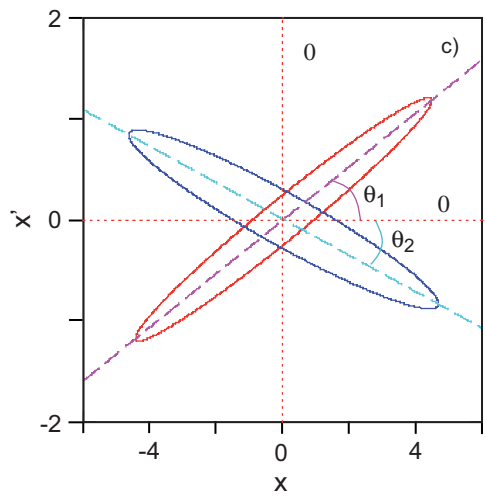

Fig. 2c

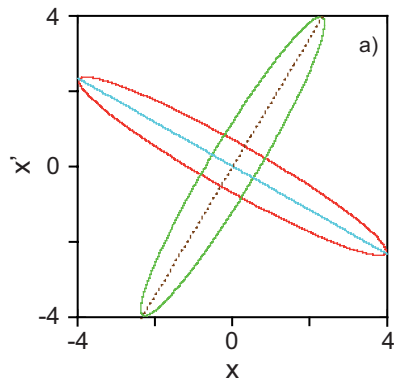

Fig. 3a

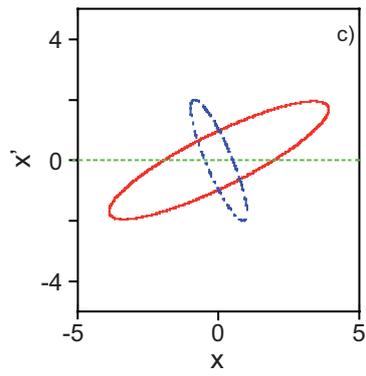

Fig. 3b

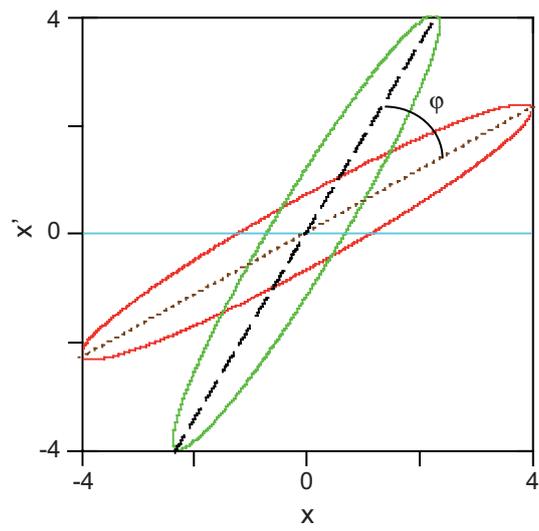

Fig. 4

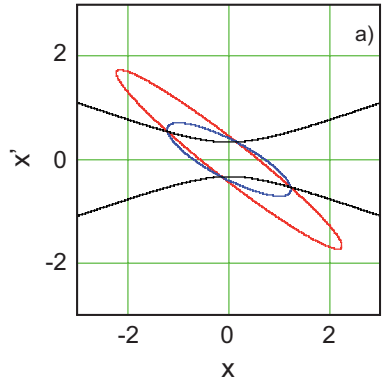

Fig. 5a

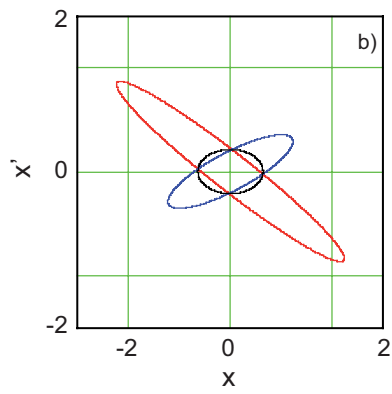

Fig. 5b

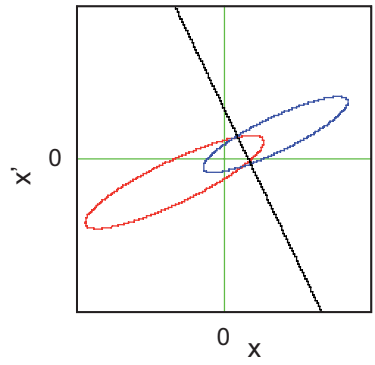

Fig. 6

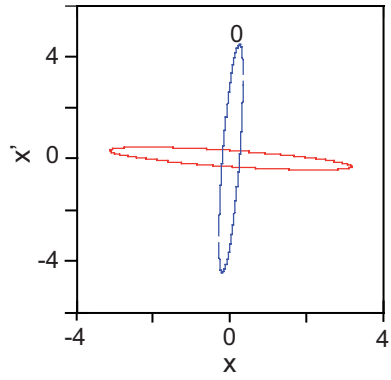

Fig. 7

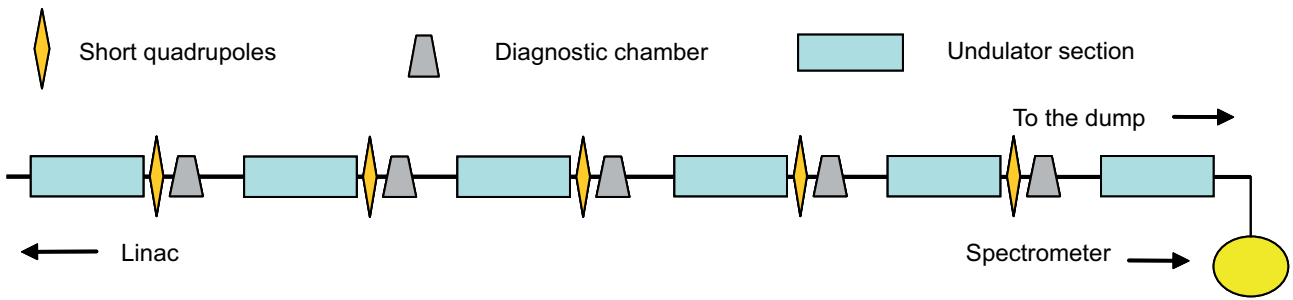

Fig. 8

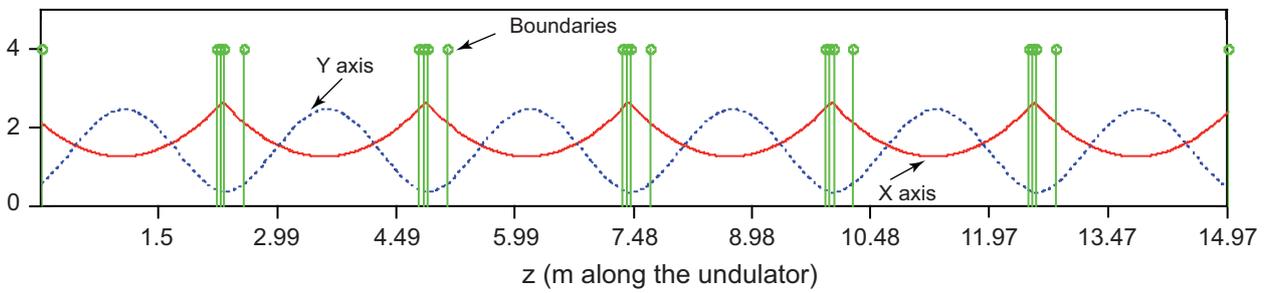

Fig. 9

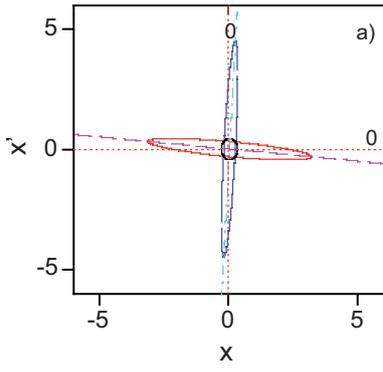
Fig. 10a
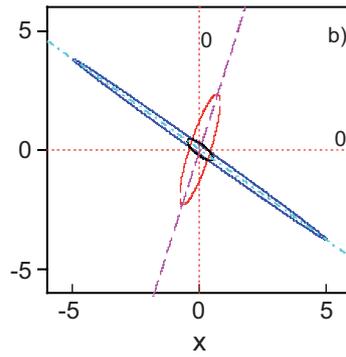
Fig. 10b
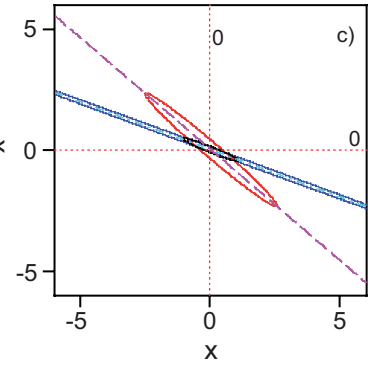
Fig. 10c
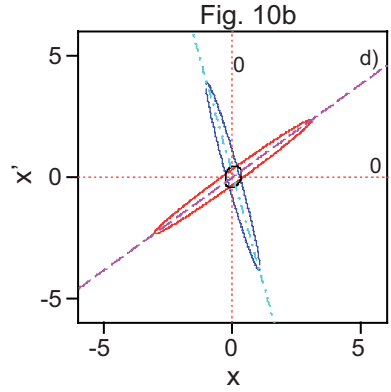
Fig. 10d

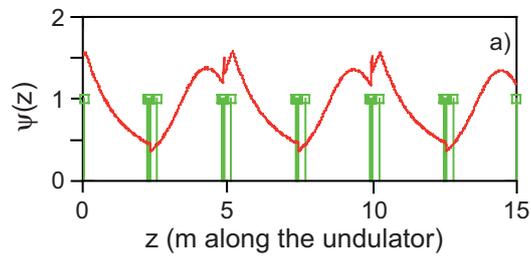
Fig. 11a

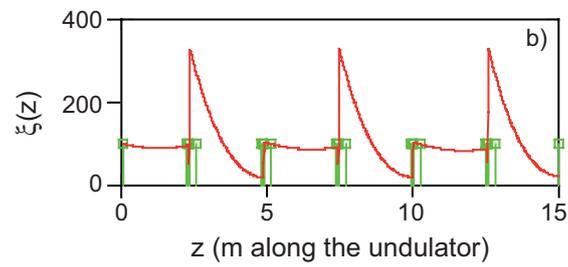
Fig. 11b

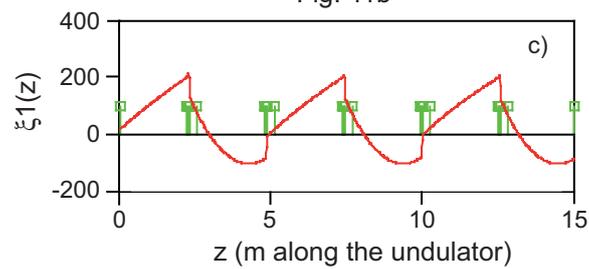
Fig. 11c

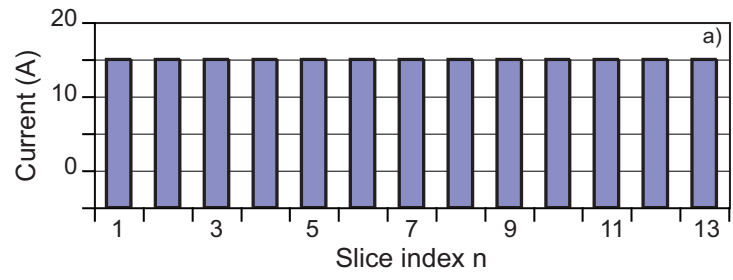

Fig. 12a

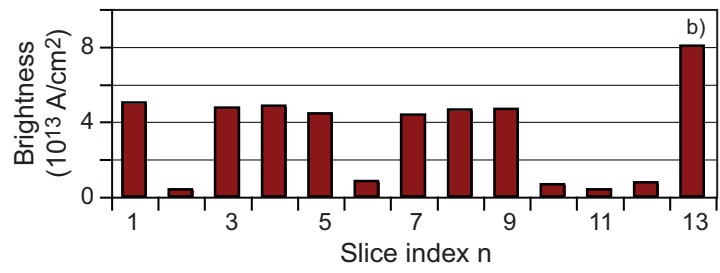

Fig. 12b

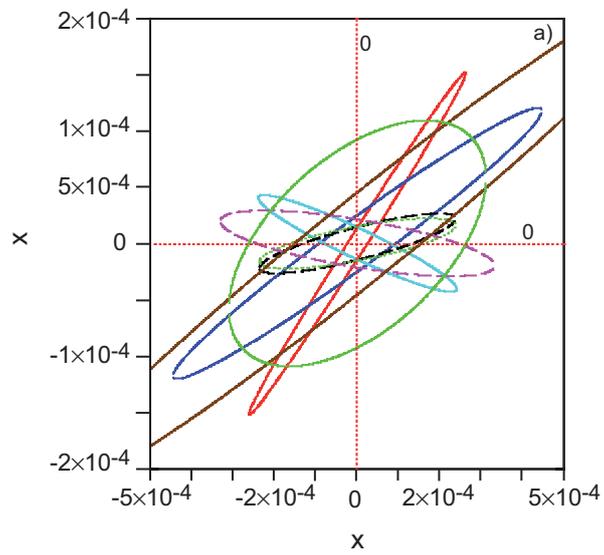

fig. 13a

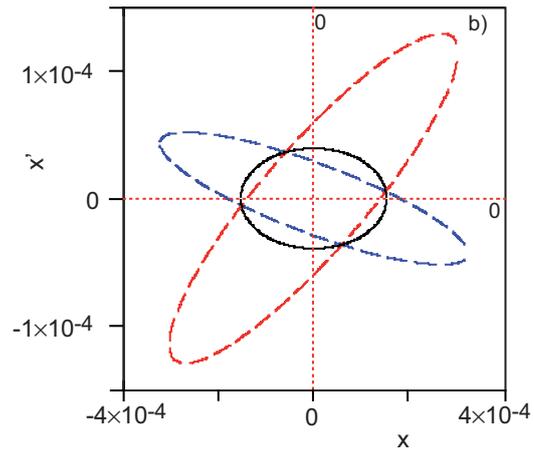

Fig. 13b

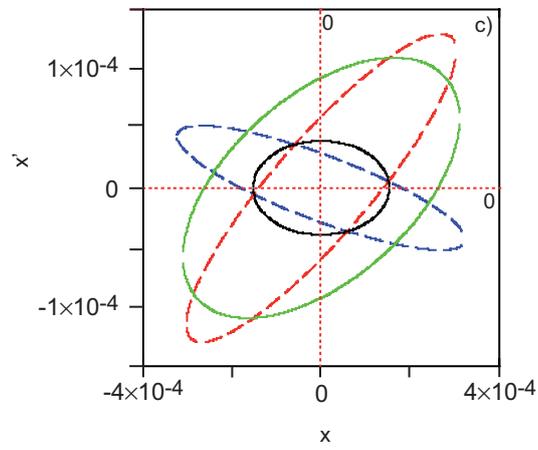

Fig. 13c

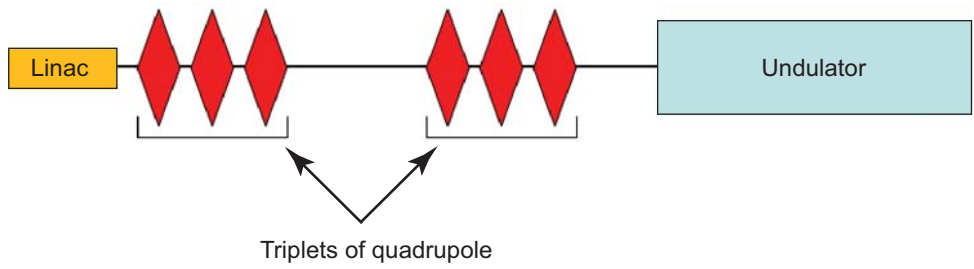

Fig. 14

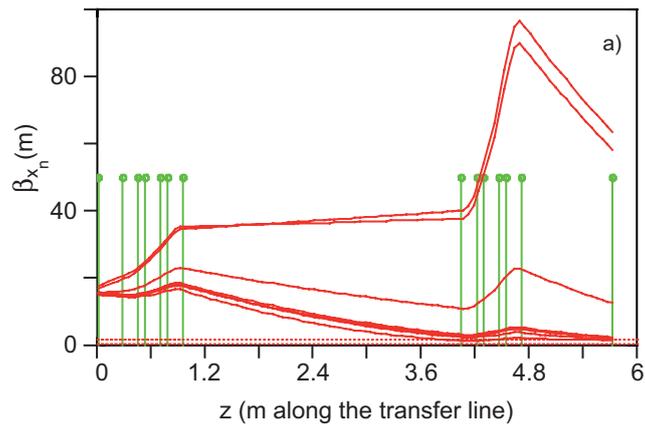

Fig. 15a

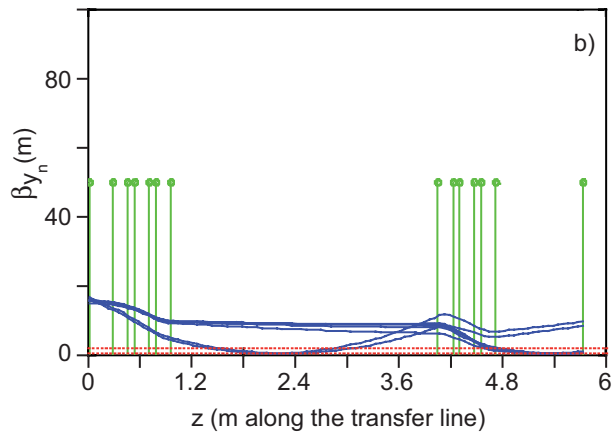

Fig. 15b

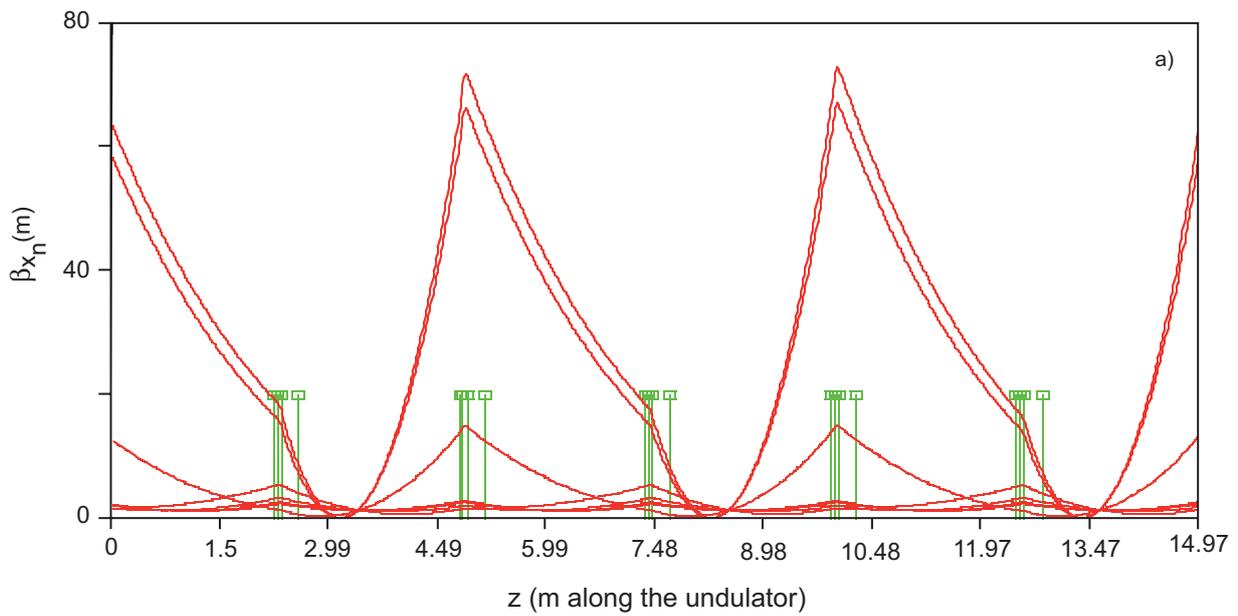

Fig. 16a

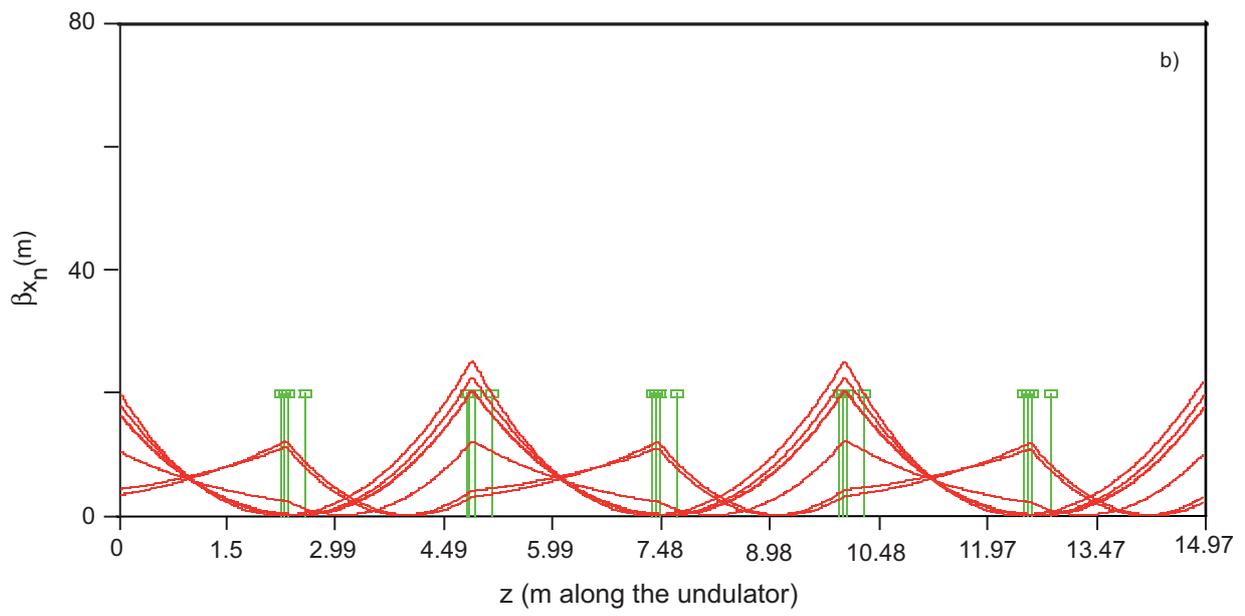

Fig. 16b

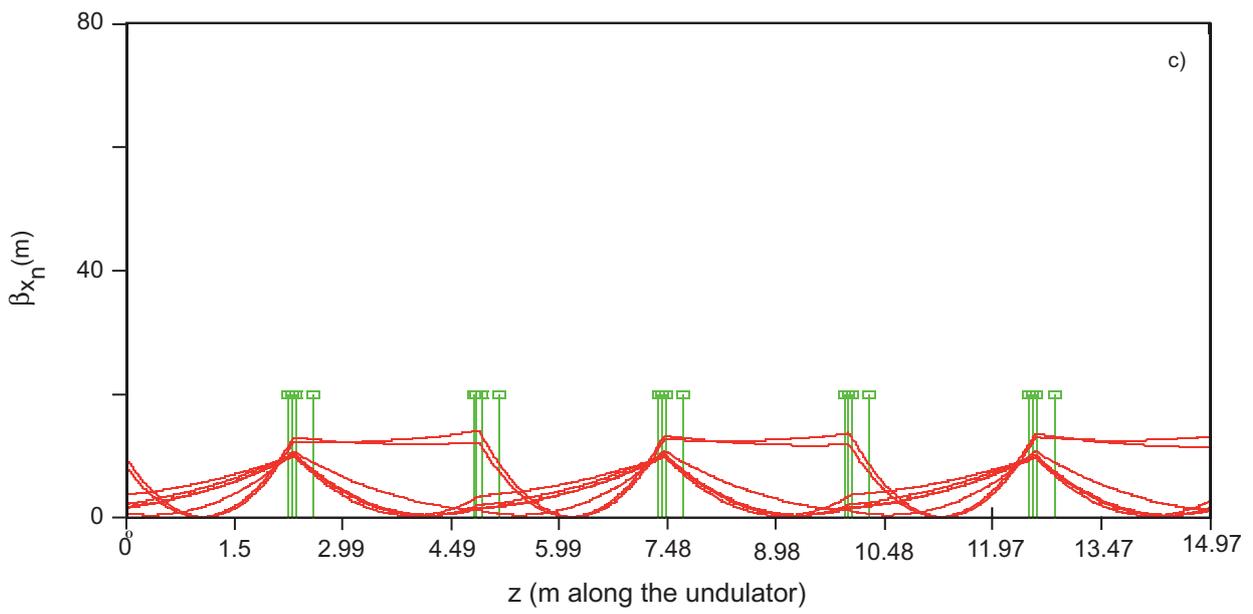

Fig. 16c

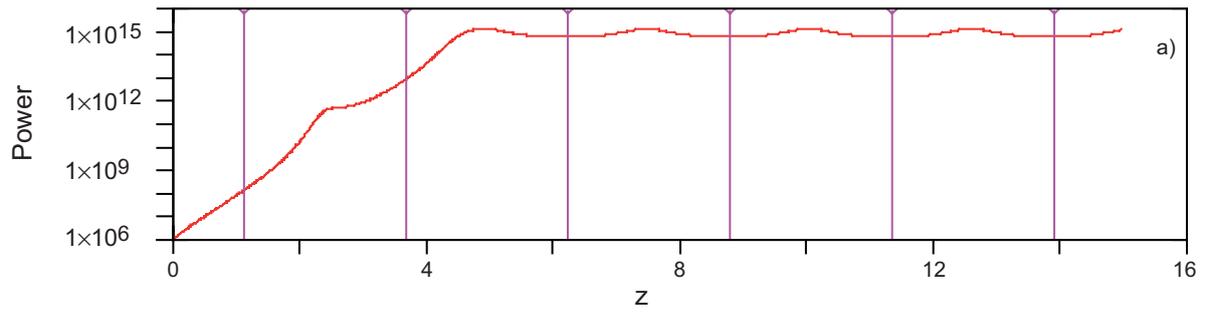

Fig. 17a

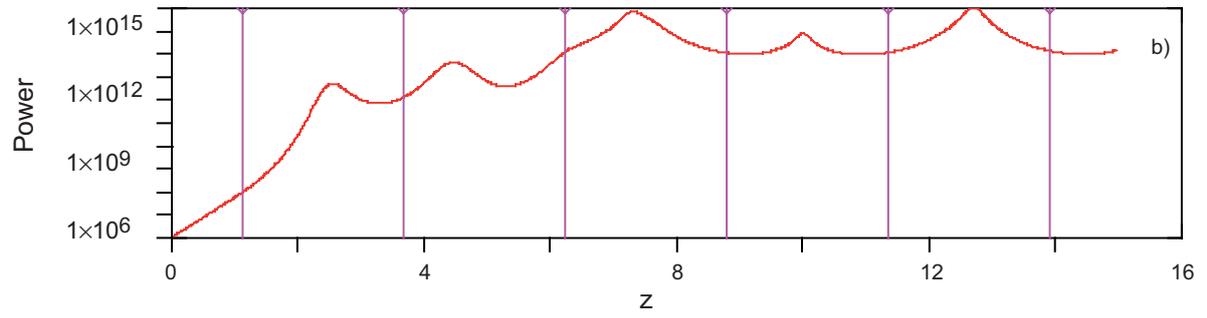

Fig. 17b

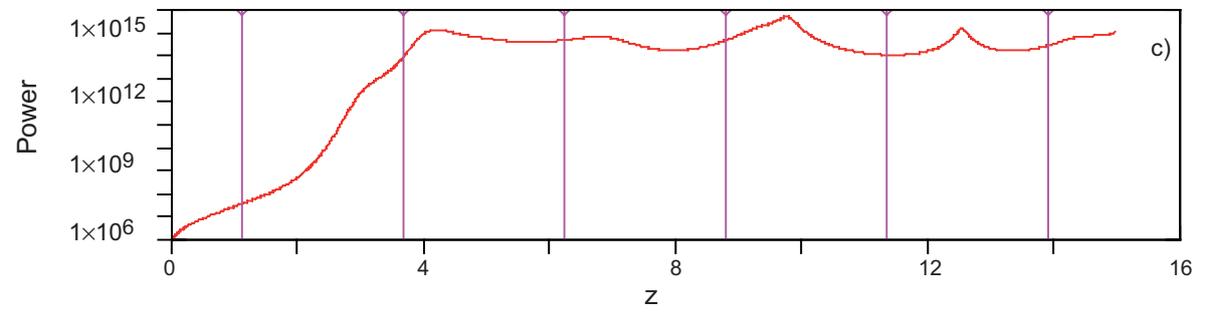

fig. 17c

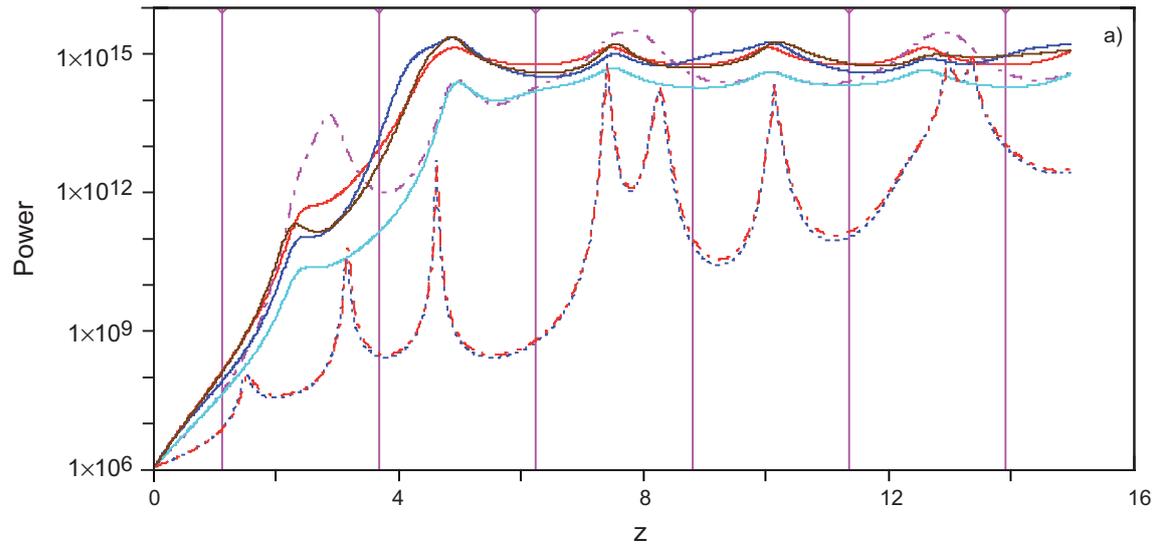

Fig. 18a

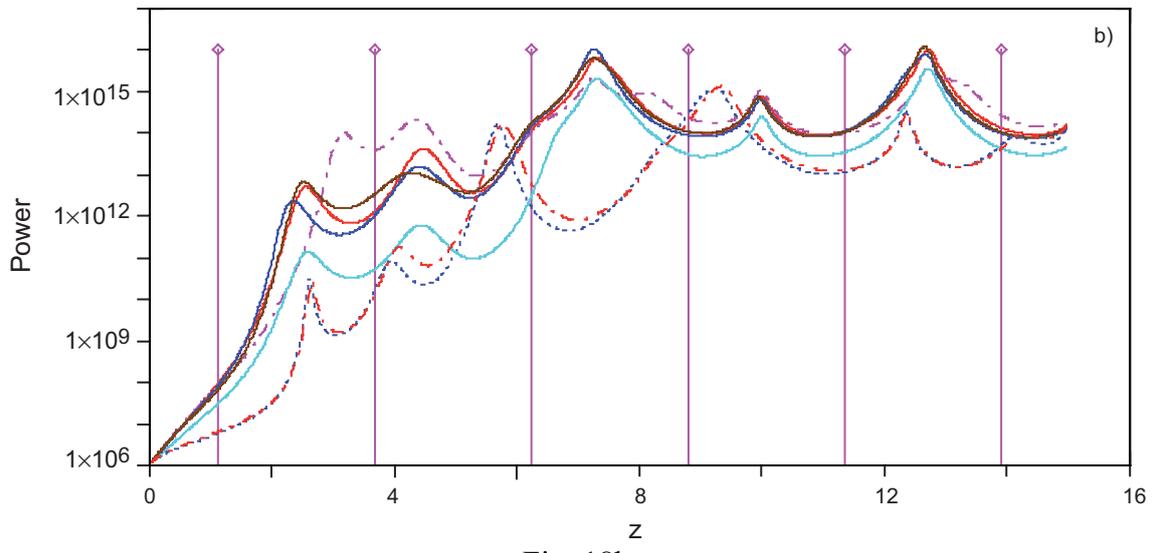

Fig. 18b

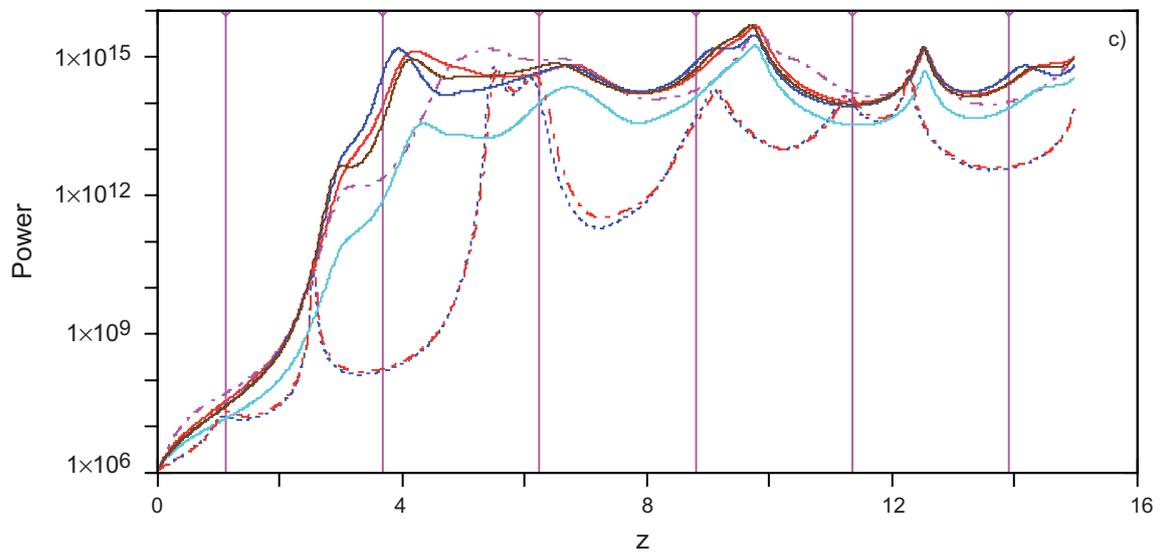

Fig. 18c

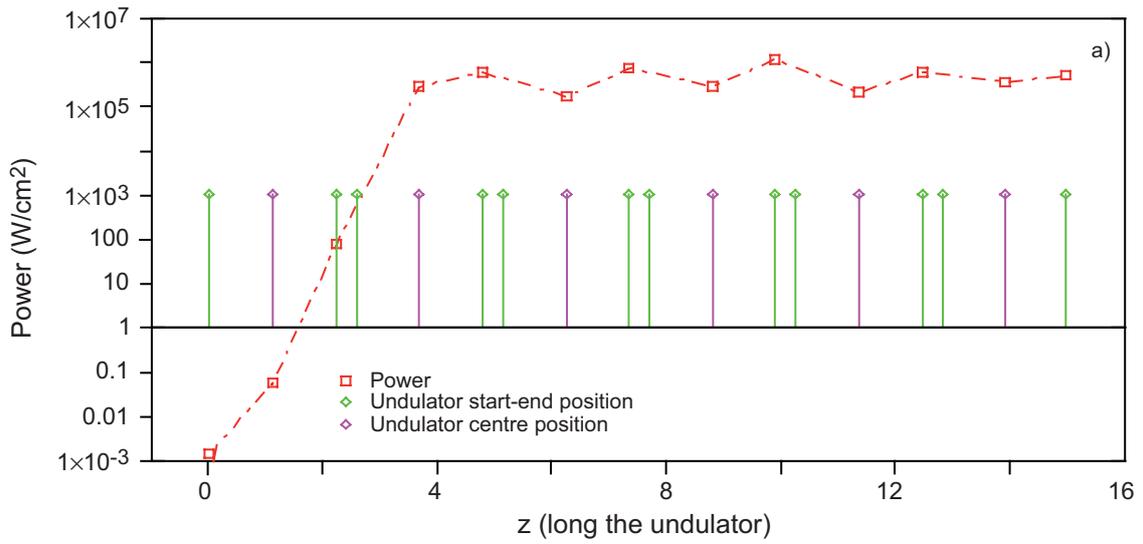

Fig. 19a

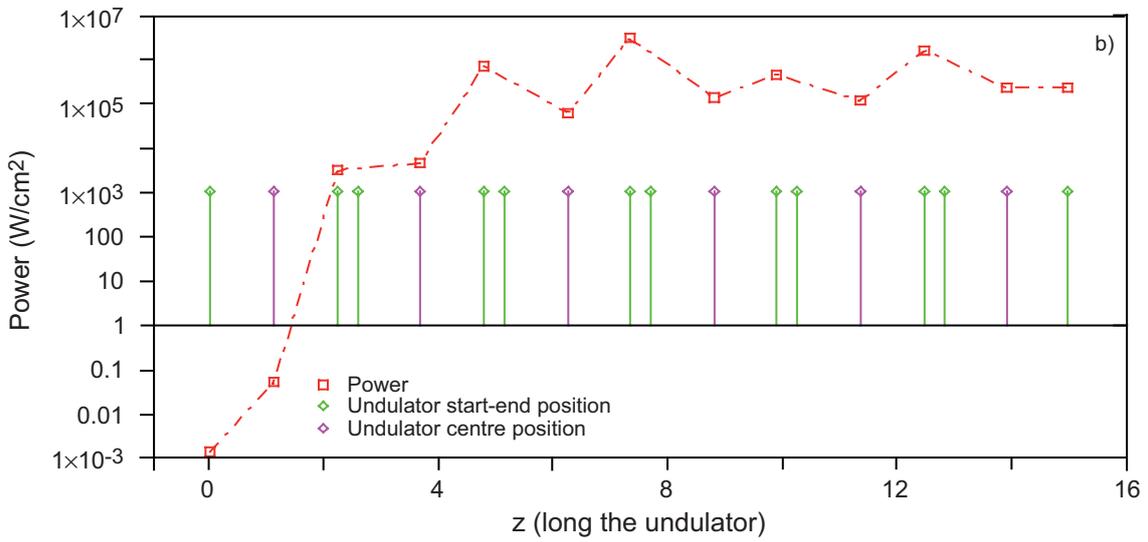

Fig. 19b

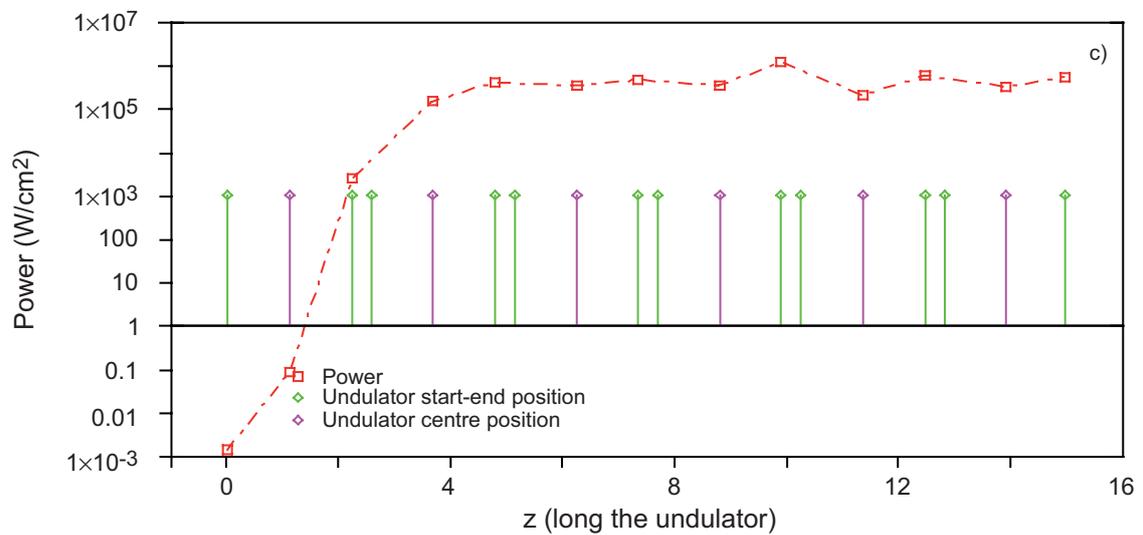

Fig. 19c

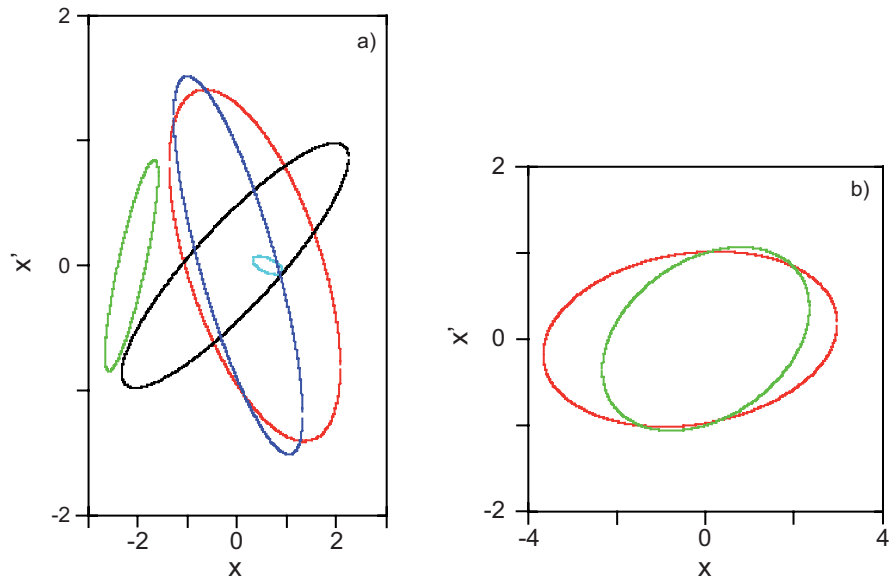

fig. 20a,b